\documentclass[aps, prl, reprint, superscriptaddress]{revtex4-1}
\usepackage[utf8]{inputenc}
\usepackage{amsmath}
\usepackage{amsfonts}
\usepackage{amssymb}
\usepackage{graphicx}
\begin{document}
\title{Quadrupolar Effects on Nuclear Spins of Neutral Arsenic Donors in Silicon}

\author{David P. Franke}
\email{david.franke@wsi.tum.de}
\author{Moritz P. D. Pfl\"uger}
\affiliation{Walter Schottky Institut and Physik-Department, Technische Universit\"at M\"unchen, Am Coulombwall 4, 
	85748 Garching, Germany}
\author{Pierre-André Mortemousque}
\altaffiliation{Present address: Institut N\'eel, 25 avenue des Martyrs, BP 166, 38042 Grenoble, France
	}
\author{Kohei M. Itoh}
\affiliation{School of Fundamental Science and Technology, Keio University, 3-14-1 Hiyoshi, 
	Kohoku-ku, Yokohama 223-8522, Japan}
\author{Martin S. Brandt}
\affiliation{Walter Schottky Institut and Physik-Department, Technische Universit\"at M\"unchen, Am Coulombwall 4, 
	85748 Garching, Germany}

\begin{abstract}
	We present electrically detected electron nuclear double resonance measurements of the nuclear spins of ionized and neutral arsenic donors in strained silicon. In addition to a reduction of the hyperfine coupling, we find significant quadrupole interactions of the nuclear spin of the neutral donors of the order of 10 kHz. By comparing these to the quadrupole shifts due to crystal fields measured for the ionized donors, we identify the effect of the additional electron on the electric field gradient at the nucleus. This extra component is expected to be caused by the coupling to electric field gradients created due to changes in the electron wavefunction under strain.
\end{abstract}

\maketitle
The electron and nuclear spins of donors in silicon have evolved as promising candidates for quantum applications, with extremely long coherence times \cite{tyryshkin_electron_2012,steger_quantum_2012,saeedi_room-temperature_2013, wolfowicz_atomic_2013}, detection down to the single-spin level \cite{pla_single-atom_2012, muhonen_storing_2014}, electrical control \cite{bradbury_stark_2006, lo_stark_2014, wolfowicz_conditional_2014,laucht_electrically_2015} and coupling to superconducting resonators \cite{zollitsch_high_2015, bienfait_controlling_2015}.
One of the challenges in the design of the involved nanostructures is the generation of uncontrolled mechanical stress due to material interfaces formed by insulators, metal gates or superconducting resonators with the Si host crystal.
These stresses are often distributed over a broad range and can be hard to predict, still their influence has to be considered for a correct quantitative description of the quantum system \cite{bienfait_reaching_2015, lo_hybrid_2015}. 

The influence of strain on the wavefunctions of donors in silicon has long been known \cite{wilson_electron_1961}. When stress is applied along the direction of two of the conduction band minima, the energy of these minima is lowered, breaking the 6-fold degeneracy of the conduction band and leading to a mixing of the electronic ground state of the donor with excited states. Because of the ensuing changes to the wavefunction, the hyperfine coupling constant $A$, which is proportional to the probability density at the nucleus $|\psi (0)|^2$, is reduced, as has been shown experimentally \cite{wilson_electron_1961,huebl_phosphorus_2006} and suggested as a possible tuning mechanism for qubits \cite{dreher_electroelastic_2011}.
In electron spin resonance (ESR) experiments at weak magnetic fields \cite{morishita_electrical_2009,bienfait_reaching_2015} or on donors with strong hyperfine coupling, such as bismuth in silicon \cite{wolfowicz_atomic_2013,bienfait_reaching_2015}, a mixing of electron and nuclear spin states leads to a significant influence of the nuclear magnetic resonance (NMR) properties on the ESR measurements \cite{wolfowicz_atomic_2013, mortemousque_hyperfine_2014}. For donors with nuclear spin $I>1/2$, therefore, the quadrupole interaction with electric field gradients has to be considered, in particular when strains are involved. While interactions with crystal field gradients cancel out in the cubic symmetry of unstrained silicon, they can significantly shift the NMR of ionized donors under strain and shear \cite{franke_interaction_2015}. For donors in their neutral charge state, the mixing with excited states under strain breaks the symmetry of the wavefunction, which could lead to an additional non-zero quadrupolar effect due to field gradients connected to the electron charge distribution \cite{mortemousque_quadrupole_2015}. NMR shifts due to these two mechanisms, crystal fields and changes to the electronic wavefunction, are not easily separated experimentally, as their influence on the observed spectra would be expected to be qualitatively equal. The necessary theoretical treatment, however, is different. While the effect of crystal fields is connected to local changes to bonds with neighboring Si atoms, the symmetry breaking of the wavefunction can be described by treating the host material as a dielectric continuum \cite{wilson_electron_1961}. In this work, we measure the NMR of neutral arsenic donors in strained silicon and identify quadrupole shifts of the resonance lines. We compare these to the shifts observed for the NMR of ionized donors in the same samples and find evidence for a component that is observed only in the neutral charge state and should be connected to changes in the wavefunction.
\begin{figure}
	\centering
	\includegraphics[width=\linewidth]{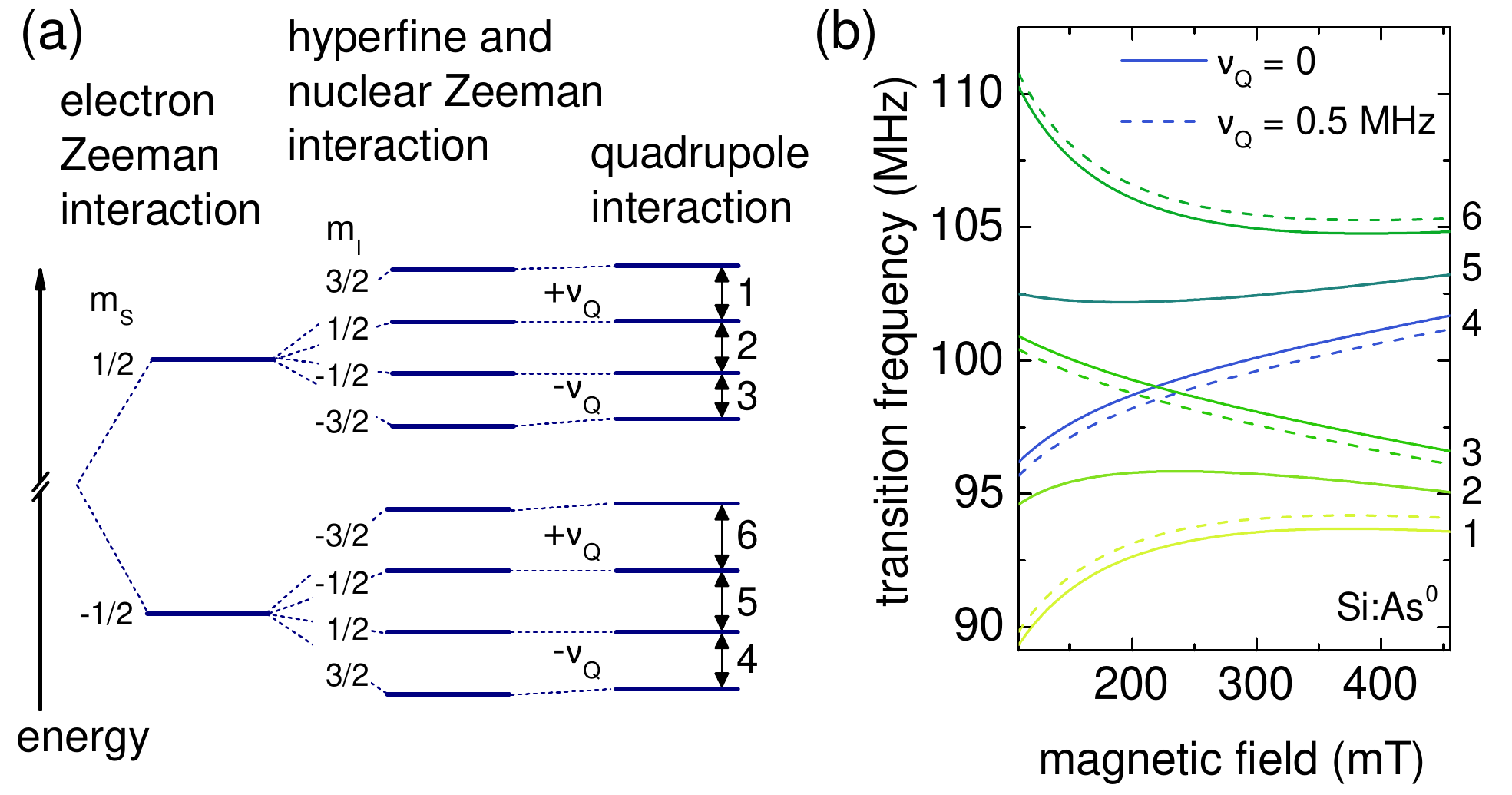}
	\caption{
		(a) Schematic representation of the energy levels of the $S=1/2$, $I=3/2$ system under the influence of different interactions in the limit of very high magnetic fields.
		(b) Dependence of the NMR transitions of neutral arsenic donors in silicon on the external magnetic field. Dashed lines show the transition frequencies considering a non-zero quadrupole interaction.}
	\label{fig:fig1}
\end{figure}

The Hamiltonian $\mathcal{H}$ characterizing neutral arsenic donors with electron spin $\mathbf{S}=\{S_x,S_y,S_z\}$ and nuclear spin $\mathbf{I}=\{I_x,I_y,I_z\}$ consists of four different interactions, here ordered by their typical strengths from highest to lowest energy
\begin{align}
	\mathcal{H}/h=f_0S_z+A\mathbf{S}\cdot\mathbf{I}-\nu_0I_z+\nu_Q\frac{1}{2}( I_z^2-\frac{5}{4})
	\mathrm{ ,}\label{eq:H}
\end{align}
where $h$ is Planck's constant. The terms describe
(i) the Zeeman interaction of the electron spin with an external magnetic field $B_z$, where $f_0=\gamma_e B_z$ with the electronic gyromagnetic ratio $\gamma_e$,
(ii) the hyperfine interaction of the electron and nuclear spins, where $A=198.35$  MHz for As in Si \cite{feher_observation_1956},
(iii) the nuclear Zeeman interaction, where $\nu_0=\gamma_n B_z$ with the nuclear gyromagnetic ratio $\gamma_n$, and
(iv) the nuclear quadrupole interaction with an effective electric field gradient $V_{33}$, here approximated to first order, with
\begin{align}
	h\nu_Q=\frac{1}{2}V_{33}eQ\cdot\frac{1}{2}(3\cos^2\vartheta -1)\mathrm{ ,}\label{eq:nuQ}
\end{align}
where $Q$ is the nuclear quadrupole moment and $e$ is the elementary charge \cite{man_nmr_2012}. The second term on the right hand side of (\ref{eq:nuQ}) describes the dependence on the angle $\vartheta$ between $B_z$ and $V_{33}$ and varies between $1$ for $\vartheta=0^{\circ}$ and $-1/2$ for $\vartheta=90^{\circ}$.

\begin{figure}
	\centering
	\includegraphics[width=\linewidth]{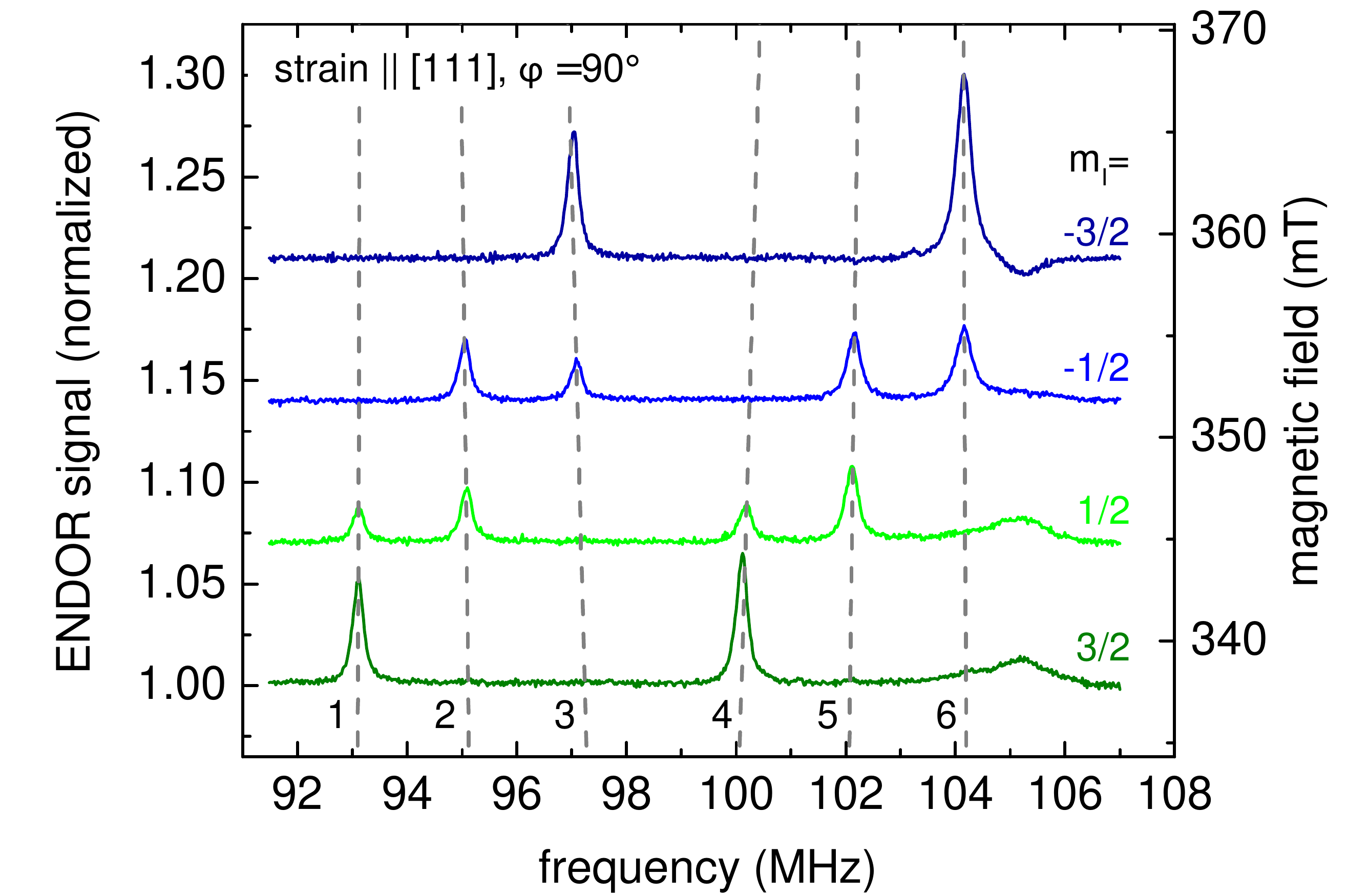}
	\caption{
		Electrically detected ENDOR spectra of As-doped silicon.  A fit to the expected magnetic field dependence  is shown as dashed lines. The broad structure at $\sim105$ MHz is caused by the frequency dependence of the radio frequency coils used in the experiment.}
	\label{fig:spectra}
\end{figure}
In a non-zero magnetic field, the eigenstates of $\mathcal{H}$ split up in two subensembles with electron spin projections $m_S=1/2$ and $-1/2$, which are further divided into four levels with different nuclear spin projections $m_I=3/2\dots -3/2$ [cf.~Fig.~\ref{fig:fig1} (a), not to scale]. The transitions which are allowed by NMR selection rules ($\Delta m_S=0$, $\Delta m_I=\pm1$) are labeled 1 through 6. In the limit of high magnetic fields and without quadrupole interactions the transition frequencies  $\nu_{m_S}=|m_S A-\nu_0|$ are equal within each subensemble. While the high-field limit is not reached in the experiments below, it still allows us to discuss the qualitative changes expected due to changes in the different coupling constants. For a reduction of the hyperfine constant $A$, we expect similar shifts towards lower frequencies for of all six transitions. On the other hand, a change in $\gamma_n$ should act on the two subensembles with different signs.
The influence of the first-oder quadrupole interaction is sketched in the last column of Fig.~\ref{fig:fig1} (a). While there is no effect on the two central transitions 2 and 5, the satellite transitions are shifted by $+\nu_Q$ (1 and 6) and $-\nu_Q$ (3 and 4).
Quantitatively this is shown in Fig.~\ref{fig:fig1} (b), where the six transition frequencies for arsenic donors are plotted as a function of the magnetic field with (dashed lines) and without (solid lines) a hypothetical quadrupole interaction of $\nu_Q=0.5$ MHz.

The samples used in this work are Czochralski-grown silicon wafers implanted with As$^+$ ions at low energies. This creates a doped region with a depth of $\sim 50$ nm below the surface. As part of the implantation damage, oxygen-vacancy complexes are formed, which, in their excited triplet state (SL1), are an efficient recombination partner for As donors \cite{franke_spin-dependent_2014,franke_spin-dependent_2014-1}. This allows us to perform electrically detected magnetic resonance (EDMR) experiments providing the high sensitivity needed. The samples remained unannealed, were contacted with Cr/Au interdigit structures and biased with typically $5$ V. Electrically detected electron nuclear double resonance (ENDOR) measurements were performed in a Bruker flexline resonator for pulsed ENDOR at a temperature of $8$ K in a He flow cryostat, illumination was provided by a red light emitting diode.
The applied measuring scheme is based on the selective ionization of donors depending on their nuclear spin state as part of a spin-dependent recombination process and is discussed in detail in Ref.~\onlinecite{franke_interaction_2015}, \onlinecite{dreher_nuclear_2012}, and \onlinecite{hoehne_submillisecond_2015}. This recombination leads to two subensembles with different charge states, ionized donors Si:As$^+$ and neutral donors Si:As$^0$. Because of the selectivity of the ionization, both ensembles are highly polarized and their NMR transitions can be accessed in the ENDOR experiments. The detection is performed on the same electron spin resonance line as the selective ionization, which means that we expect a positive signal for neutral nuclear spin resonances and a negative signal for resonances of ionized donors \cite{hoehne_submillisecond_2015}.

\begin{figure}
	\centering
	\includegraphics[width=\linewidth]{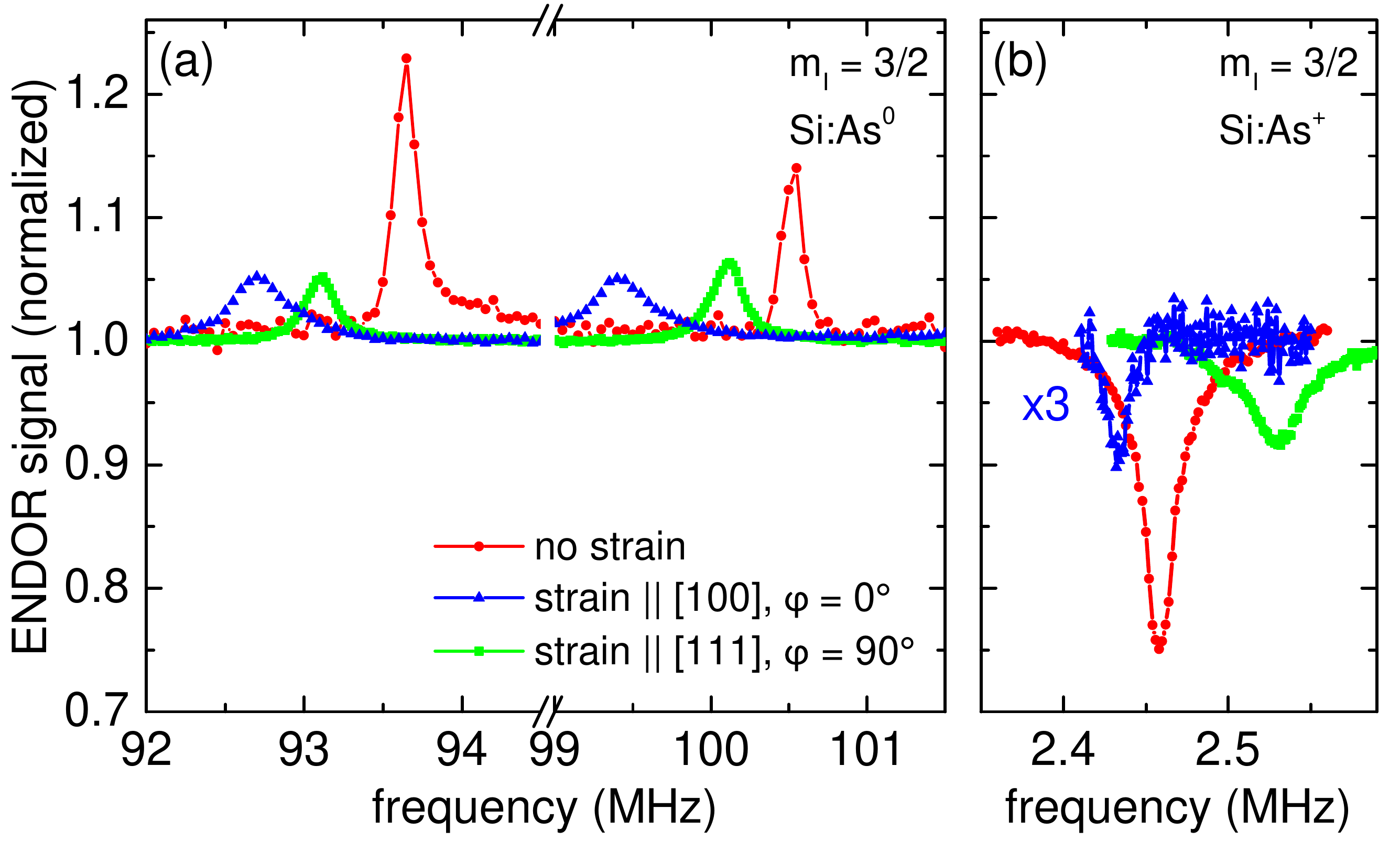}
	\caption{
		Comparison of ENDOR spectra of samples with and without strain. (a) Neutral donors As$^0$. (b) Ionized donors As$^+$.
	}
	\label{fig:strain}
\end{figure}

Figure \ref{fig:spectra} shows electrically detected ENDOR spectra recorded on the four hyperfine-split electron resonance lines. As expected, the nuclear magnetic resonances of the neutral As donors are observed as an enhancement in signal amplitude. Each measurement is sensitive to transitions which involve the nuclear spin state chosen for ionization and read-out (given next to the traces in Fig.~\ref{fig:spectra}), hence every resonance is detected in two of the spectra. The observed line positions are well described by a fit to the expected magnetic field dependence of the spin system Hamiltonian (\ref{eq:H}) which is shown as dashed lines.

To study the effect of strain on these resonances, samples with different crystal orientations were cemented onto sapphire substrates, which at low temperatures induces strain due to the different thermal expansion coefficients \cite{ibach_thermal_1969, lucht_precise_2003}. We assume the resulting strain to be uniaxial and normal to the Si/sapphire interface. Note that the angle $\varphi$ between this normal and the magnetic field $B_z$ is not per se equal to $\vartheta$, which describes the orientation of the generated electric field gradient. Figure \ref{fig:strain} (a) shows spectra obtained for Si/sapphire stacks with samples from [100] and [111] wafers. Compared to the line position of an unstrained sample, both As$^0$ related resonances in each of the spectra are shifted towards lower frequencies. This indicates that the dominant effect on the line position is a reduction of the hyperfine coupling to the donor electron and not a quadrupole interaction, which would shift the two lines of each spectrum in different directions (cf.~Fig.~\ref{fig:fig1}). Such a reduction of the hyperfine constant $A$ is expected for strain that lifts the symmetry of the six conduction band minima in silicon, which is expected for our [100] sample. More quantitatively, $A(\chi)/A(0)=\frac{1}{2} \left((\frac{\chi}{6}+1)/(\frac{\chi^2}{4}+\frac{\chi}{3}+1)^{1/2}+1\right)$ with the unitless valleystrain $\chi$ (cf.~Ref.~\onlinecite{wilson_electron_1961}), which is a measure for the strain-induced asymmetric change of the electron wavefunction. This leads to asymmetric electric field gradients, which interact with the nuclear spin of the donor via quadrupole interaction. We therefore expect that any quadrupole shift due to the electron wavefunction is connected to $\chi$.
For strain in [111] direction, the energy of all conduction band minima is lowered by the same amount and $\chi=0$ \cite{wilson_electron_1961}. Still, we also observe a reduced hyperfine coupling in the [111] sample, suggesting that the strain in the Si/sapphire stack is not purely along the [111] axis and $\chi\neq 0$ also in this case.

\begin{table*}%
	\begin{tabular}{lcccccccc}
		strain& $\varphi$ &$A$ & $\chi$ & $\gamma_n/h$ & $\nu_Q^0$ & $\nu_Q^+$ & $\Delta\nu_Q$\\
		\hline
		& & MHz & &kHz/mT&kHz&kHz&kHz\\
		\hline
		\hline
		no strain & 90$^{\circ}$& 198.25(5) & -0.09(7) & 7.30(10) & - & -&-\\
		\hline
		strain $\parallel [100]$ & 0$^{\circ}$ & 196.20(2) &-0.42(1)& 7.28(10) & 65(8) & 34(5) & 31(11)\\
		& 90$^{\circ}$& 196.39(3) & -0.40(1)&7.25(10) &-19(15) & -5(5) & -14(18)\\
		\hline
		strain $\parallel [111]$ & 0$^{\circ}$& 197.27(2) & -0.30(1)&7.28(5) & 149(5) & 128(5) & 21 (9)\\
		& 90$^{\circ}$& 197.32(2) &-0.29(1)&7.27(5) & -75(5) & -63(5) & -12 (9)
		
	\end{tabular}
	\caption{Summary of the fit results of the neutral and ionized NMR measurements on three different samples. Estimated experimental errors are given in parentheses. For the unstrained sample, the fit was performed without considering the quadrupole term in the Hamiltonian since it did not significantly improve the fit.}
	\label{tab:tab1}
\end{table*}

At lower frequencies [Fig.~\ref{fig:strain} (b)], the spin resonance of the ionized donors As$^+$ is observed as a negative signal. Since $A=0$ in this case, the observed As$^+$ NMR frequency $\nu_+$ of the $m_I=3/2\leftrightarrow 1/2$ transition shown here is given by $\nu^+=\nu_0-\nu_Q^+$ with the quadrupole interaction frequency $\nu_Q^+$ of the ionized donor. For the strained samples, shifts of $\nu_Q^+=34(5)$ and $-63(5)$ kHz are observed for the [100] and [111] samples, respectively, in agreement with previous measurements \cite{franke_interaction_2015} and the expected angular dependence for $\vartheta=\varphi$.

\begin{figure}
	\centering
	\includegraphics[width=\linewidth]{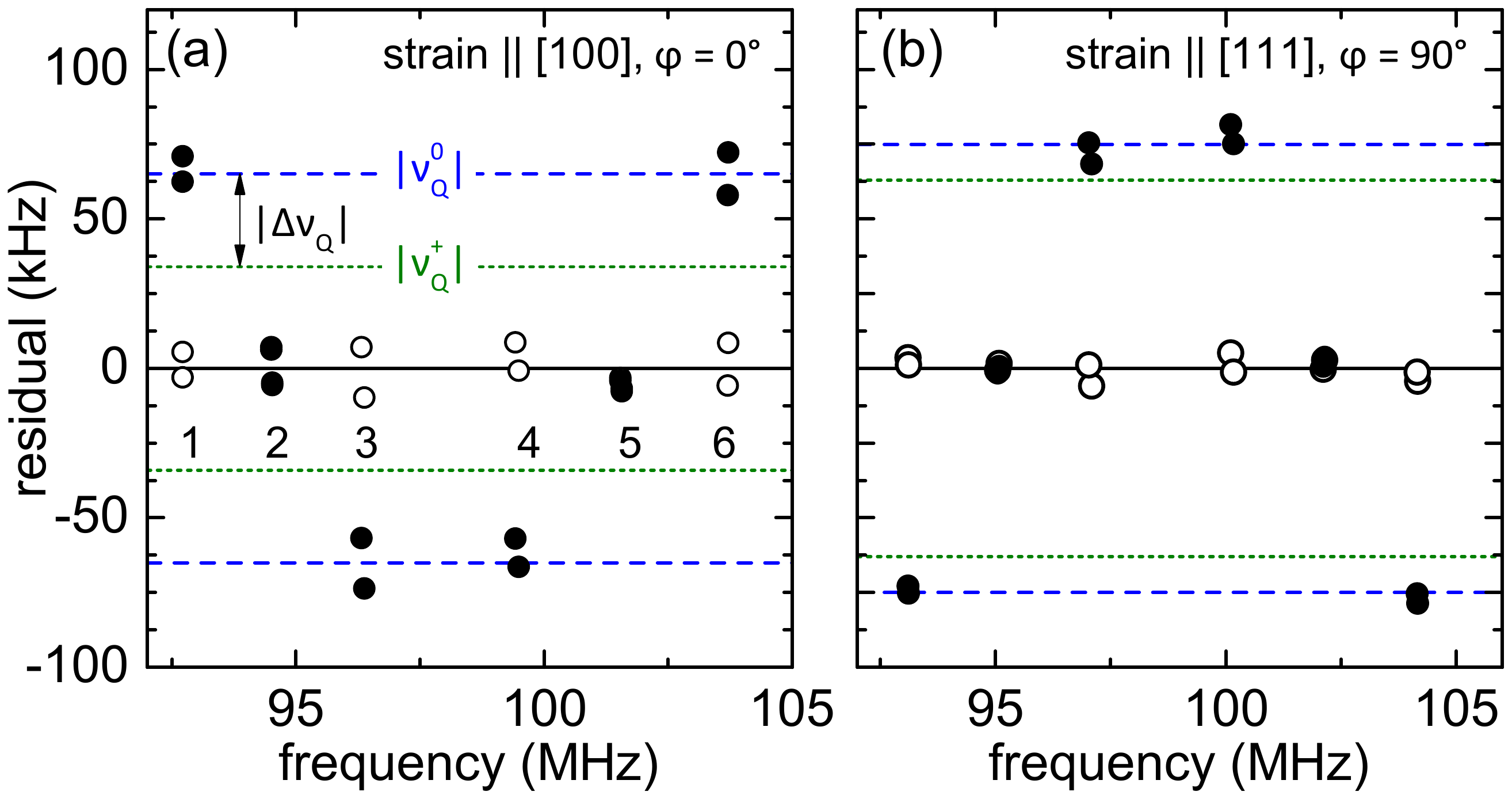}
	\caption{
		Residuals showing the deviation from a theoretical description by hyperfine interaction and nuclear Zeeman interaction only (full circles) and when also considering a quadrupole interaction $\nu_Q^0$ (open cicles). $+\nu_Q^0$ and $-\nu_Q^0$ are shown as blue dashed lines, the green dotted lines represents the shift $\nu_Q^+$ determined in measurements of the ionized donors Si:As$^+$.
	}
	\label{fig:residuals}
\end{figure}

To give a more detailed analysis of the observed resonances of the neutral donors, the peak positions are extracted by fitting with pseudo-Voigt functions (shape factors $\sim 0.5$). For the resulting data, least square fits to the expected peak positions are performed, using the hyperfine constant $A$ and the nuclear gyromagnetic ratio $\gamma_n$ as fitting parameters. We then plot the deviation of the observed peak position from these fits to study any behavior that can not well be described considering only these two interactions. The resulting residuals for the As$^0$ NMR in the sample strained along the [100] axis are shown as full circles in Fig.~\ref{fig:residuals} (a) for $\varphi=0$. Clearly, a systematic deviation from 0 is observed, which is positive and about equal for resonances 1 and 6, negative but of similar absolute value for 3 and 4, and close to zero for resonances 2 and 5. This mirrors the expected shift due to a quadrupole interaction (cf.~Fig.~\ref{fig:fig1}). When including the term describing the quadrupole interaction in the fit, a significantly better result is achieved (empty circles) and the remaining deviations seem unsystematic and are expected to reflect the experimental noise. The resulting quadrupole shift $\nu_Q^0$ is shown as a blue dashed line in Fig.~\ref{fig:residuals}. When the sample is rotated in the magnetic field by $\Delta\varphi=90^{\circ}$, the sign of $\nu_Q$ changes (data not shown). Performing the same analysis for the data obtained on the [111] Si:As sample for $\varphi=0^{\circ}$ (data not shown) and $\varphi=90^{\circ}$ [Fig.~\ref{fig:residuals} (b)], a similar systematic deviation is observed. Again, the residuals are very well explained by a quadrupolar effect and the measurements at different angles are very well described by the expected angular dependence (\ref{eq:nuQ}) for an effective electric field gradient generated along the strain axis ($\vartheta=\varphi$). The constants determined by these fits are summarized in Tab.~\ref{tab:tab1}. 

To investigate the origin of these quadrupole shifts, we compare them to the shifts $\nu_Q^+$ observed on the ionized donors Si:As$^+$ for each of the samples, which are shown as green dotted lines in Fig.~\ref{fig:residuals} (a) and (b). For both samples, these shifts are slightly smaller than those observed for the neutral donors $\nu_Q^0$ (blue lines). This suggests that we observe an additional quadrupole interaction connected to the donor electron, the strength of which can be estimated by taking the differences $\Delta\nu_Q = \nu_Q^0-\nu_Q^+$ of the frequency shifts. The values for both samples are given in Tab.~\ref{tab:tab1}. For the [100] sample,  $\Delta\nu_Q$ of 31(11) and -14(18) kHz are observed at $\varphi= 0$ and 90 degrees, respectively. The factor $\sim-1/2$ between the two measurements is once more in agreement with the expected angular dependence (\ref{eq:nuQ}), if the electric field gradient $V_{33}^{\Delta}$ connected to the electron wavefunction is generated parallel to the strain axis in the [100] direction of our sample. Since the conduction band minima in Si are along the (100) axes, this would indeed be expected. 
Using (\ref{eq:nuQ}), we can calculate an electric field gradient \mbox{$V_{33}^{\Delta}=7.8\pm10\times 10^{18}$ V/m$^2$}, which is generated upon the application of a uniaxial strain $\epsilon_{\perp}=2.3(5)\times10^{-4}$ along the [100] axis, as calculated from the observed change of the hyperfine coupling \cite{huebl_phosphorus_2006}. For the [111] sample, $\Delta\nu_Q$ of 21(9) and -12(9) kHz are observed at $\varphi= 0$ and 90 degrees, respectively. As discussed above, while in principle no change to the symmetry of the wavefunction is expected for strain in [111] direction, the reduction of $A$ indicates the presence of a (weaker) valley strain $\chi$, probably due to a slightly different strain axis in our samples. Consequently, a weaker but non-zero $\Delta\nu_Q$ would be expected in this sample as well, which is in agreement with our measurements. The observed shifts correspond to an effective electric field gradient at the nucleus \mbox{$V_{33}^{\Delta}=6.1\pm5\times 10^{18}$ V/m$^2$} which is generated approximately along the strain axis. Comparing the values for $\chi$ and $V_{33}^{\Delta}$ measured in the two samples, the observed monotonous and possibly linear dependence strongly suggests that strain components altering the wavefunction symmetry are responsible for the observed additional quadrupolar effects in the neutral charge state.

In summary, we report shifts to the NMR of neutral donors in strained silicon which agree with the behavior expected for quadrupolar effects. These shifts differ from the quadrupole shifts due to crystal field gradients which we determine via the NMR of ionized donors and indicate the presence of an additional quadrupole interaction of a similar order of magnitude. This interaction, which is only observed in the neutral charge state, is likely to be connected to electric field gradients caused by changes to the wavefunction of the donor electron. 
The measured shifts are, however, smaller than those due to the strain-induced change to the hyperfine interaction by more than one order of magnitude, clearly demonstrating that the quadrupole interaction is not the dominant mechanism in high-field NMR experiments on arsenic donors silicon. For ESR measurements on mixed spin states, the influence of $A$ and $\nu_Q$ on the resonance positions strongly depends on the observed transition and applied magnetic field \cite{mortemousque_quadrupole_2015}. In particular, the influence of the hyperfine interaction vanishes at certain fields \cite{mortemousque_hyperfine_2014}, so that $\nu_Q$ could be the central influence of strain on the spectrum in this case. Even though the errors to the values measured here are still large, we can conclude that such wavefunction-induced gradients are of the same order of magnitude as the effects due to crystal field gradients and that both effects should be included in the modeling of strain effects on arsenic donors in silicon. For a more complete understanding of the dependence of $\Delta\nu_Q$ on the wavefunction, similar measurements on other group-V donors would be desirable. This should also help to identify the involved anti-shielding parameters, which could lead to much stronger effective field gradients in heavier donors \cite{kaufmann_electric_1979}. Furthermore, a more precise determination of the resulting quadrupole shifts could possibly be achieved when the inhomogeneous broadening of the NMR of the neutral donors is reduced. To this end, unixaial stress would have to be applied in a very homogeneous way. In addition, broadening due to superhyperfine interactions with surrounding $^{29}$Si nuclei could be reduced in isotopically controlled Si \cite{itoh_isotope_2014}.
Still, since the relevant strains will be smaller in most nanostructures than in our Si/sapphire stacks, our results give a valuable estimation of the magnitude of quadrupolar interactions of nuclear spins with deformations of the electron wavefunction, which should motivate further modeling and can provide a test for future theories.

\begin{acknowledgments}
The authors would like to thank Hans-Werner Becker for the implantation and Manabu Otsuka for sample characterization. The work at TUM was supported financially by DFG via SFB 631 and SPP 1601, the work at Keio by KAKENHI (S) No. 26220602 and JSPS Core-to-Core.
\end{acknowledgments}

\bibliography{bib4}
\end{document}